\documentclass[%
twocolumn,
showpacs,
amsmath,amssymb,
aps,
prb,
floatfix,10pt
]{revtex4-1}
\usepackage{amsmath}
\usepackage{graphicx}
\usepackage{amssymb}
\usepackage{color}
\usepackage{ulem}

\newcommand{\comm}[1]{\ensuremath{\left[#1\right]_-}}

\newcommand{\fn}{\ensuremath{\mathbf{n}}}
\newcommand{\fs}{\ensuremath{\mathbf{s}}}
\newcommand{\fJ}{\ensuremath{\mathbf{J}}}
\newcommand{\dsd}{\ensuremath{\Delta_{\text{sd}}}}

\newcommand{\fJp}{\ensuremath{\mathbf{J}_p}}

\newcommand{\fey}{\ensuremath{\mathbf e_{y}}}

\newcommand{\half}{\ensuremath{\frac{1}{2}}}

\newcommand{\mean}[1]{\ensuremath{\left\langle#1\right\rangle}}

\newcommand{\dx}{\ensuremath{\partial_x}}
\newcommand{\dt}{\ensuremath{\partial_t}}

\newcommand{\bi}{\ensuremath{\beta^{(i)}}}
\newcommand{\bii}{\ensuremath{\beta^{(ii)}}}
\newcommand{\ar}{\ensuremath{\alpha_R}}

\begin{document}
\title{Nonequilibrium Rashba field driven domain wall motion in ferromagnetic
nanowires}

\author{Martin Stier,$^1$ Reinhold Egger,$^2$ and Michael Thorwart$^1$}
\affiliation{$^1$I. Institut f\"ur Theoretische Physik, Universit\"at Hamburg, 
Jungiusstra{\ss}e 9, 20355 Hamburg, Germany\\ 
$^2$Institut f\"ur Theoretische Physik, 
Heinrich-Heine-Universt\"at D\"usseldorf, 40225 D\"usseldorf, Germany}

\begin{abstract}
We study the effects of spin-orbit interaction (SOI) on the
current-induced motion of a magnetic (Bloch) domain wall 
in ultrathin ferromagnetic nanowires.
The conspiracy of spin relaxation and SOI is shown to generate a
strong nonequilibrium Rashba field, which can dominate even for 
weak SOI.  This field causes intricate spin precession and a 
transition from translatory to oscillatory wall dynamics with 
increasing SOI.  We show that current pulses of
different lengths can be used to efficiently control the domain wall motion. 
\end{abstract}


\maketitle

\section{Introduction}

The efficient and reliable manipulation of magnetic 
microstructures forms the basis for most information storage devices
used nowadays.  Commonly, an applied local magnetic field
controls the magnetization in a given domain. The alignment
direction encodes a classical bit, and the data storage density is 
limited by the domain size. Recent advances in nanofabrication 
have implemented Berger's proposal \cite{Berger78} of moving a domain wall (DW)
by a current-induced spin torque \cite{Review}.
This may allow for ultrasmall magnetic devices with 
high data storage density in ferromagnets carrying a 
spin-polarized current. The local magnetization can be electrically
controlled by a spin torque arising from the exchange coupling 
 of the local spins to conduction electron spins.  
Across a wide DW, the polarization of the itinerant spins 
adiabatically follows the magnetization direction,
angular momentum is transferred due to total spin conservation, 
and DW motion along the current direction is induced.  In addition to this
adiabatic spin torque \cite{Review}, a nonadiabatic spin torque plays a
prominent role \cite{nonad1,gerrit,mt2007,physrep,garate}:  
This so-called ``$\beta$-term'' is due to 
spin relaxation, which causes the itinerant spin polarization 
to ``lag'' behind the local magnetization. 
This nonequilibrium contribution to the torque critically determines the DW
velocity and shape, as well as the depinning and critical (Walker breakdown) 
current densities.

Ferromagnetic nanowires are natural candidates for building ultrafast memory
and logic devices that rely on nanoscale current-induced DW
motion.  While quite high DW velocities ($\approx 100$~m$/$s)
are possible in permalloy (NiFe) nanowires \cite{hayashi,meier},
these setups often suffer from limited reproducibility, strong DW pinning, and
low critical currents.  Following earlier proposals \cite{Obata,Manchon}, 
recent experiments \cite{Miron11a} realized Co nanowires  
in an AlO$_x/$Co$/$Pt trilayer structure,  where
structural inversion asymmetry causes an
interfacial electric field and thus a Rashba spin-orbit 
interaction (SOI) \cite{Matos}.
This electronic SOI is strong, can be tuned by electrostatic gating, 
and allows one to largely circumvent the above 
problems \cite{Miron11a,Miron10,Miron11b}.  
The observed DW velocities of up to $400$ m/s (and other interesting
features, e.g., DW motion against the current 
direction) were attributed \cite{Miron11a} to a
conspiracy of the $\beta$-term and a field-like adiabatic Rashba spin torque
\cite{Obata,Manchon,Ryu11,Hayashi12,Ryu12,Martinez12,tsutsui},
denoted ${\bf T}_1$ below.  While ${\bf T}_1$ does not involve spin 
transfer and can be traced back to the electronic bandstructure, it 
depends on the current and can switch the magnetization.  
These exciting experimental observations and their technological 
promise have triggered further theoretical works 
\cite{Kim12,Wang12a,Pesin12,Duine12,Wang12}, which
draw a more complex picture, involving also a
Sloncezwski-type nonadiabatic Rashba spin torque
(denoted ${\bf T}_2$ below)  due to the interplay of SOI 
and spin relaxation occurring under nonequilibrium conditions.

Given the complexity of this problem, we here aim to understand 
current-induced DW motion in a ferromagnetic Rashba nanowire
in the simpler one-dimensional (1D) limit. This limit
allows for the analytical calculation of the full 
current-induced nonadiabatic spin torque appearing in
the Landau-Lifshitz-Gilbert (LLG) equation for the space- and time-dependent
magnetization profile.  Previous experiments \cite{Miron11a} have 
used Co nanowires of diameter $\approx 500$~nm, much thicker
than the few-channel nanowires studied here, but future experiments 
could approach this ultrathin-wire limit.  By numerical solution of the 
LLG equation including the full spin torque, with the nanowire initially 
containing a Bloch DW, we predict ultrafast DW velocities in current-pulsed
setups.  Surprisingly, we find that already weak Rashba spin-orbit couplings
have a huge effect on the DW dynamics due the appearance of a ``nonequilibrium
Rashba field.''  Spin-orbit coupled ferromagnetic nanowires are 
thus predicted to allow for ultrafast and efficient DW dynamics.    

\section{Model and LLG equation}
We here
study an ultrathin ferromagnetic Rashba nanowire (along the $x$-direction), 
where localized spins create the magnetization profile
$-M_s \fn (x,t)$, with unit vector $\fn$ and saturation magnetization $M_s$.
The magnetization dynamics is governed by the LLG equation \cite{Review}, 
\begin{equation} \label{eq_llg}
 \dt\fn=-\fn\times \mathbf H_{\text{eff}}+\alpha\fn\times\dt\fn+\mathbf T,
\end{equation}
where the effective field $\mathbf H_{\text{eff}}(x)$
 generates magnetic texture (for a specific DW profile, see below) 
in the absence of a current, and the Gilbert damping
 parameter $\alpha$ depends on 
intrinsic material properties. The spin torque,
$\mathbf T(x,t)$, encapsulates all current-induced contributions due
to the exchange interaction between localized moments
and itinerant electrons, described here within the standard $sd$ model.
The conduction electrons carry a spin-polarized electric current, and
also experience the Rashba SOI.  
Remarkably, in the 1D limit, analytical results for the 
full spin torque can be obtained by employing the textbook 
Sugawara representation \cite{Gogolin}. We discuss the main steps 
of the derivation next (we often use units with $\hbar=1$).
Technical details can be found in the Appendix \ref{app_deriv}. 

\section{Derivation}

At low energy scales, 
itinerant electrons in the nanowire have 
linear momentum $k_x\approx pk_F$, 
where $p=+$\ $(p=-)$ stands for a right (left) mover and 
$k_F$ is the Fermi momentum.  Away from the band bottom, the dispersion
relation can be linearized, where the Fermi velocity $v$ sets the slope.
  The electronic spin density vector,
${\bf s}(x)={\bf J}_R+{\bf J}_L$, and the 
corresponding spin current, ${\bf J}(x)=v({\bf J}_R-{\bf J}_L)$, 
are thereby expressed in terms of 
chiral spin density vectors ${\bf J}_{R,L}(x)$
for $p$-movers only.  Similarly, the
(scalar) density, $\rho_c=\rho_R+\rho_L$, and the charge current flowing 
through the nanowire, $I_c= ev (\rho_R-\rho_L)$, follow from the
chiral particle densities $\rho_{R,L}$.  
This separation into left- and right-moving parts
is typical for 1D systems and allows for
the analytical progress reported here.  With the exchange coupling $\dsd$,
the spin torque in Eq.~\eqref{eq_llg} is  
$\mathbf T (x,t) = -\dsd\fn\times\langle\bf{s}\rangle$,
where the average is over the electronic degrees of freedom 
taking into account the SOI.  Using the dimensionless
Rashba coupling $\alpha_R$, the single-particle Hamiltonian
receives the contribution
$(\dsd\alpha^{}_R/2k_F)[\sigma_y k_x -\sigma_x k_y]$ \cite{winkler},
where $\sigma_{x,y}$ are spin Pauli matrices
and the channel-mixing term $\propto k_y$
is negligible for ultrathin nanowires \cite{starykh}.
As discussed below, ${\bf T}$ follows by solving the 
Heisenberg equations of motion for ${\bf J}_{p}$ (with $p=+,-=R,L$),
\begin{eqnarray} \label{eqheom}
(\dt + pv\dx) \fJp & = & - \dsd \fJp\times (\fn+p\alpha_R\mathbf e_y)\\
\nonumber &-& \sum_{\nu=i,ii}\beta^{(\nu)}\dsd \left(\fJp-
\fJp^{(\nu)}\right ) ,
\end{eqnarray} 
where spin relaxation (the last term) 
has been included phenomenologically within the
relaxation-time approximation; $\mathbf e_y$ is the unit vector in the
$y$-direction.  We identify two competing relaxation 
mechanisms characterized by different stationary configurations, 
${\bf J}_p^{(i)}$ and ${\bf J}_p^{(ii)}$, 
and different rates governing the relaxation into them,
$\beta^{(i)} \dsd$ and $\beta^{(ii)}\dsd$, respectively.
Mechanism (i) arises due to the contact of the
nanowire to ferromagnetic leads, which inject the spin-polarized
current $I_s=\hbar PI_c/2e$
with spin polarization factor $-1\le P \le 1$ \cite{brat}.  
Fluctuations then try to establish the stationary distribution 
${\bf{ J}}_{p}^{(i)} = \frac12 P \rho_p \fn$ \cite{balents}.
Mechanism (ii) instead describes spin relaxation to the
intrinsic stationary solution of Eq.~(\ref{eqheom}),
${\bf{J}}_p^{(ii)}(x) = \frac12 P \rho_p [\fn(x)+p\alpha_R \mathbf e_y]$.

For a given magnetization profile $\fn(x,t)$,
since the typical space-time variations of $\fn$ are slow 
compared to electronic variations, 
Eq.~(\ref{eqheom}) can now be solved analytically by an
iterative gradient expansion \cite{mt2007}.  In this approach, 
we first determine the spin torque under
the assumption that all space-time derivatives $\partial_{x,t}\fn$ 
can be discarded. The resulting zeroth-order result, 
${\bf T}={\bf T}^{(0)}(x,t)$, is then used to obtain the 
first-order correction to the spin torque, ${\bf T}^{(1)}$,
calculated by retaining terms $\sim\partial_{x,t} {\bf n}$ 
but omitting all higher derivatives. We consider only those two
orders and, for clarity, we write down only the leading terms in
$\alpha_R\ll 1$ below.  Complete spin torque expressions, 
valid for arbitrary $\alpha_R$, are lengthy and given in the Appendix \ref{app_deriv}.
Our numerical results for the DW dynamics (see below) were obtained 
with the complete expressions but remain very similar
when using only the small-$\alpha_R$ spin torque.

\section{Current-induced Rashba spin torque}

We recover the two 
Rashba spin torques discussed in the Introduction already
from the zeroth-order result, ${\bf T}^{(0)}={\bf T}_1+{\bf T}_2$.
The field-like spin torque, ${\bf T}_1=-\fn\times{\bf H}_R$,
can be written in terms of a ``nonequilibrium Rashba field,''
\begin{equation}\label{t1}
{\bf H}_R= H_R^{(0)} \mathbf e_y, \quad
 H_R^{(0)}=\alpha_R I_s\dsd/v ,
\end{equation}
while the Sloncezwksi-type nonadiabatic Rashba spin torque reads
\begin{equation}\label{t2}
\mathbf T_2 = (\beta^{(i)} \alpha_R I_s\dsd/v)\
 \fn\times(\fn\times\mathbf e_y)  . 
\end{equation}
The structure of both spin torques agrees with previous results 
\cite{Kim12,Wang12a,Pesin12,Duine12,Wang12}.
For the first-order correction ${\bf T}^{(1)}$, we have
\begin{equation} \label{t11}
\mathbf T^{(1)}= - I_s \partial_x\fn +\beta^{(i)} I_s \fn 
\times \partial_x\fn -H^{(1)}_R\fn\times {\bf e}_y,
\end{equation}
where additional time-dependent terms $\sim\partial_t \fn$ (not shown)
 effectively renormalize the Gilbert damping parameter. 
In Eq.~\eqref{t11}, we recognize the adiabatic spin torque (first term)
as well as the ``$\beta$-term'' (second term), where both are present already 
without SOI. Finite $\alpha_R$ then renormalizes the prefactors in both
terms (see Appendix \ref{app_deriv}).  The last term in Eq.~(\ref{t11}) 
contributes to the nonequilibrium Rashba field, ${\bf H}_R
= H_R \mathbf e_y$ with $H_R=H_R^{(0)}+H_R^{(1)}$, where we find 
\begin{equation} \label{hr}
H_R = \frac{\alpha_R I_s \dsd}{v} -
\frac{\alpha_R^2\beta^{(ii)} I_s}{(\beta^{(i)}+\beta^{(ii)})^2} 
\dx n_y + {\cal O}(\alpha_R^3).
\end{equation}
The first term comes from Eq.~\eqref{t1} and has been discussed before
\cite{Obata,Manchon,Ryu11,Hayashi12,Ryu12,Martinez12,tsutsui}.
The second term in Eq.~\eqref{hr} is new and may
dominate for $\beta^{(i,ii)}\ll\alpha_R^2$. This implies that a 
strong nonequilibrium Rashba field already emerges
even for rather weak Rashba couplings. 
While nonadiabatic torque contributions 
[$\mathbf T_2$ in Eq.~\eqref{t2} and the second term in Eq.~\eqref{t11}] 
are due to relaxation mechanism (i),  $H_R^{(1)}$  is mainly caused by
mechanism (ii)\cite{HR1}.

\begin{figure}[t!]
\centering
\includegraphics[width=.99\linewidth]{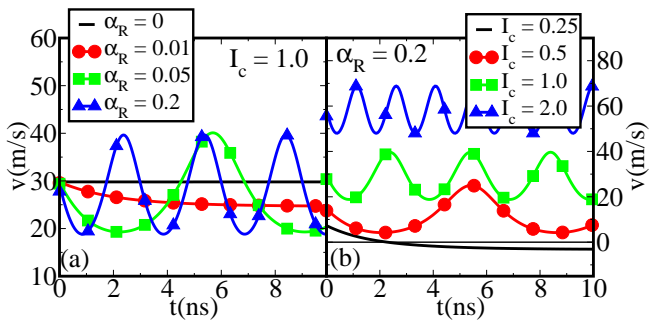}
\caption{\label{f1} 
(Color online) Velocity of the DW center vs propagation time for
(a) different $\alpha_R$  and (b) different  $I_c~(\rm{10^{12}A/m^2})$. 
An increase of either $\alpha_R$ or $I_c$ increases the Rashba field $H_R$. 
At the critical value $H^{\rm WB}$, we find
a current-induced Walker breakdown and the velocity starts to oscillate. 
Further increase of $\alpha_R$ or $I_c$ increases the oscillation
frequency.}
\end{figure}

\section{Numerical simulation}

To obtain explicit results for 
the current-induced DW motion, we have performed 
numerical simulations. We
study a Bloch-$y$ DW created by the effective magnetic field
$\mathbf H_{\text{eff}}=J\partial_x^2\fn+Kn_y\mathbf e_y-
K_{\perp}n_x\mathbf e_x$ in
Eq.~(\ref{eq_llg}), where $J$ is an exchange coupling between 
localized moments, $K$ and $K_{\perp}$ are anisotropy constants.
Measuring length (time) in units of $x_0=0.5$~nm 
 ($t_0=1$~ps), we adopt the following parameter values:  $J=5.2$, 
$K=0.185$, $K_{\perp}=0.008$, $\alpha=\beta^{(i)}=\beta^{(ii)}=0.06$. 
This choice corresponds to Ta/CoFeB/MgO, a material with comparably 
strong SOI as in Co nanowires and well characterized 
parameter values \cite{Hayashi12}. Furthermore, unless stated otherwise,
we set $\dsd/v=0.2, v=2\times 10^5~$m$/$s, 
$\alpha_R=0.2, P=1, \rho_c=I_c/v,$ and $I_s=0.06$. 
The latter value implies a charge current density $I_c=10^{12}$~A$/$m$^2$. 
We have verified from additional calculations for other 
parameter sets (not shown) that the results below are generic.

\subsection{Steady-state current}

Numerical solution of Eq.~\eqref{eq_llg} 
with a time-independent current density $I_c$ confirms 
that the DW motion is strongly influenced by the
Rashba SOI, both concerning the instantaneous (momentary) motion
and the asymptotic long-time behavior. 
Let us first discuss the time evolution of the  
momentary velocity, see Fig.~\ref{f1}. With
increasing SOI strength $\alpha_R$, 
we observe a transition from purely translatory motion,
with (asymptotically) constant DW velocity,
to a regime with superimposed oscillations in the time-dependent
 velocity.  Our numerical analysis reveals that the oscillations 
stem mainly from the Rashba field $\mathbf H_R$. In fact, the DW 
magnetization $\fn$ is found to precess around the $y$-axis,
which is the direction of the Rashba field (cf. Appendix). 
This behavior strongly resembles a \textit{field}-induced
Walker breakdown, where the oscillations appear 
once $H_R$ exceeds a certain critical Walker field $H^{\rm
WB}$. 
As is known from the field-induced case, this
critical field depends on the perpendicular anisotropy, $H^{\rm WB}\propto
K_{\perp}$ \cite{Thomas2007}.  Within the regime $H_R>H^{\rm WB}$, we 
observe that the oscillation amplitude increases with $K_\perp$, but
remains independent of all other parameters, while
the oscillation frequency is $\propto H_R$.
Since $H_R$ is affected by a spin-polarized current $I_s$ and not 
only by $\alpha_R$, the DW motion can be effectively controlled by $I_s$, 
see Fig.~\ref{f1}(b). The impact of all other parameters on the DW motion is
less pronounced and not discussed here.

\begin{figure}[tb]
 \centering
\includegraphics[width=.85\linewidth]{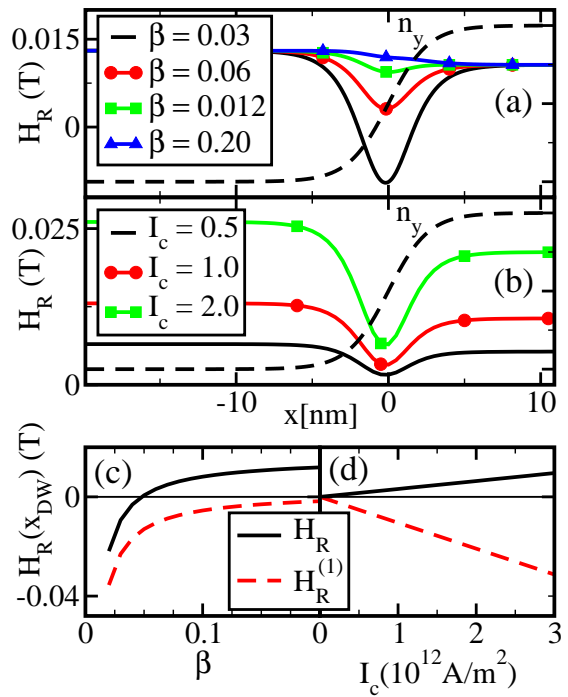}
\caption{(Color online) \label{f2}
Momentary Rashba field $H_R$ (in Tesla) close 
to the DW center (taken at $x_{DW} = 0$). We put 
$\alpha=\beta^{(i)} =\beta^{(ii)}\equiv\beta$ 
and show the behavior for (a) different $\beta$ (with $I_c=10^{12}$~A$/$m$^2$),
and (b) for different current densities 
$I_c$ (in $10^{12}$A$/$m$^2$), with $\beta=0.06$.  For comparison, the
$n_y$-component of the magnetization is shown as dashed curve. 
The last two panels illustrate the Rashba field $H_R(x_{DW})$
and the contribution $H_R^{(1)}$ directly
at the DW center: (c) $H_R$ vs $\beta$,  and 
(d) $H_R$ vs $I_c$.  }
\end{figure}

\subsection{Nonequilibrium Rashba field}

The spatial distribution of
the momentary Rashba field, $H_R(x)$, is illustrated in Fig.~\ref{f2}.  
We focus on the most relevant behavior of this field close to
the DW center.  Near the DW center, $\partial_x n_y$  
is sizable, and we then find that $H_R^{(1)}$ dominates
as long as the damping parameter $\beta$ stays small,
in accordance with Eq.~(\ref{hr}).  Far away from the DW center,
as expected, $H_R^{(1)}$ plays no significant role.
Similarly, for large $\beta$, Fig.~\ref{f2}(c) demonstrates that the 
contribution $H_R^{(1)}$ is suppressed against $H_R^{(0)}$, cf.~Eq.~(\ref{t1}).
For sufficiently small but finite $\beta$, we find that $H_R^{(1)}$ 
always provides the dominant contribution
to the nonequilibrium Rashba field near the DW center.

\begin{figure}[t!]
 \centering
\includegraphics[width=.7\linewidth]{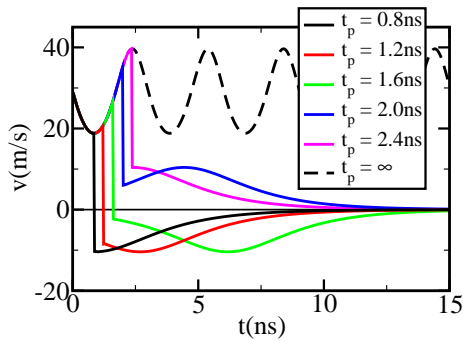}
\caption{ \label{f3}
(Color online) Velocity of the DW center vs propagation time for five  
current pulses of different length $t_p$. After the pulse, 
the DW drifts for a certain time in either positive or negative direction 
(depending on the oscillation phase at the pulse end). 
The dashed curve shows the corresponding result
for a steady-state current.}
\end{figure}

\subsection{Domain wall response to current pulses}

Experiments 
are often carried out with current pulses instead of steady-state
currents \cite{Miron11a,Vogel2012}. 
For rectangular current pulses of duration $t_p$, we find that
current-induced DW motion closely resembles the behavior in the field-driven
case above the Walker breakdown \cite{Review}.  Here,
the DW does not immediately stop at the end of
the pulse, but instead \textit{drifts} for a certain time with 
nearly constant velocity, see Fig.~\ref{f3}.  
Interestingly, the drift direction depends on the
actual phase of the oscillation at the end of the pulse. Therefore, 
even if the DW initially moves forward due to the current pulse, it may still 
end up in a backward position relative to its starting point. For
short current pulses, the drifting mechanism can 
completely dominate the total DW displacement.

\begin{figure}[t!]
 \centering
\includegraphics[width=.85\linewidth]{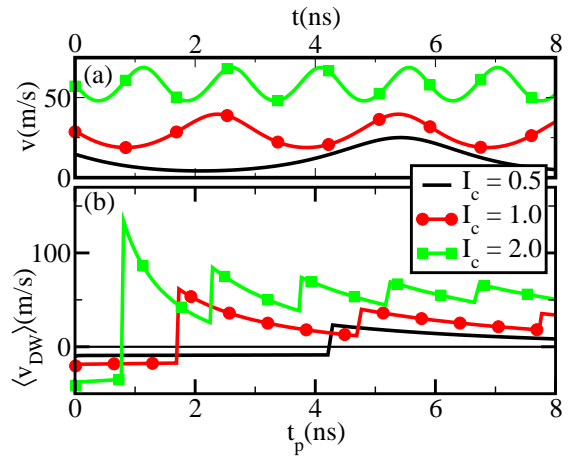}
\caption{\label{f4}
(Color online) (a) Momentary velocity of the DW center vs propagation
time for a
steady-state current.  (b) Average DW velocity
$\mean{v_{DW}}=x_{DW}(t\to\infty)/t_p$ vs current pulse 
duration $t_p$ with $I_c~(10^{12}\rm{A/m^2})$.  Depending on the momentary velocity at 
the pulse end, the DW drifts in different directions. 
Especially for short pulses, this may strongly affect $\mean{v_{DW}}$. }
\end{figure}

\subsection{Global average wall velocity}

In the current-pulsed case, a 
straightforward way to determine the average DW velocity 
is to measure the initial and final DW
position before and directly after the pulse, and then to divide this 
distance by the pulse duration $t_p$ \cite{Thomas2007}. 
In experiments, however, it is often difficult 
to read out the DW position right at the end of the pulse, and 
usually one obtains the final position only somewhat later. 
The extracted average velocity thus
coincides with the real one only if the DW stops instantly when the
pulse ends.  Figure \ref{f3} demonstrates that such assumptions are not valid
in general: The final position, $x_{DW}(t\to\infty)$, typically
deviates strongly from the location right after the pulse, $x_{DW}(t_p)$.
In the following, we study the experimentally more accessible 
``global'' average DW velocity, $\mean{v_{DW}}=x_{DW}(t\to\infty)/t_p$. 
A major issue determining $\mean{v_{DW}}$ is the oscillation phase
reached by the DW at time $t=t_p$. 
With changing $t_p$, the oscillatory DW magnetization
ends up in different configurations associated with different 
drift directions and final positions, 
$x_{DW}(t\to\infty)=x_{DW}(t_p)+x_{\text{drift}}$.
While $x_{DW}(t_p)\sim t_p$ is determined by the pulse
duration,  $x_{\text{drift}}$ can have either sign.
For very long pulses ($t_p\to\infty$), we have $|x_{DW}(t_p)|\gg
|x_{\text{drift}}|$, and the conventional average velocity, $x(t_p)/t_p$, 
coincides with $\mean{v_{DW}}$.  For short pulses, however, 
they may differ considerably.  We find that 
the $t_p$-dependence of the global average velocity $\mean{v_{DW}}$ 
exhibits a sawtooth-like behavior,  see Fig.~\ref{f4}. 

\begin{figure}[tb]
 \centering
\includegraphics[width=.85\linewidth]{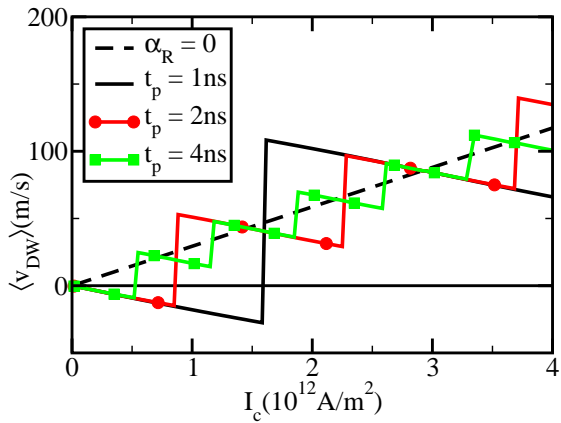}
\caption{\label{f5}
(Color online) Global average velocity of the DW center 
$\mean{v_{DW}}$ vs 
current density $I_c$ for fixed pulse duration $t_p$. 
Since $I_c$ determines the oscillation frequency, the 
pulse can end up in different oscillation states, 
causing positive or negative drift.  This can strongly change 
$\mean{v_{DW}}$ and causes DW motion against the current direction 
for small $I_c$.}
\end{figure}

Tuning the pulse duration $t_p$ is not the only way to control the motion
of the DW. For fixed $t_p$, we can also change the oscillation
frequency of the DW motion by changing the current density $I_c$, see 
Fig.~\ref{f5}.  We again find a sawtooth-like dependence of 
$\mean{v_{DW}}$ on $I_c$,
which roughly oscillates around the velocity found for $\alpha_R=0$.
For small values of $I_c$, the DW even moves against the current flow. 
Due to the strong $I_c$-dependence of $\mean{v_{DW}}$, 
small changes in the current can here lead to completely different
types of DW motion.  

\section{Conclusions}

In this paper, we have shown 
that  Rashba spin-orbit interactions can qualitatively affect 
the current-driven DW motion in ferromagnetic nanowires.  
Their main influence is encoded in a nonequilibrium Rashba field $H_R$,
which is responsible for magnetization precession and thereby 
for qualitative changes in the DW dynamics.  Remarkably, 
we find that the conspiracy of spin relaxation and spin-orbit coupling
is able to generate a dominant contribution to $H_R$ 
even for weak Rashba couplings.  As a consequence,
above a critical current-induced Rashba field $H_R>H^{\rm WB}$, 
the DW dynamics is predicted to exhibit
oscillatory features  (similar to the field-driven case 
above the Walker breakdown) and a nontrivial dependence of the
DW velocity on the current density. In fact, the DW motion can 
turn from a purely translationary into an oscillatory dynamics
due to precession of the DW magnetization around the local
nonequilibrium Rashba field. 
When using pulsed currents with variable 
pulse duration, even the direction of the DW motion can be controlled.
All predicted effects should be observable with present-day
experimental techniques.

\section*{Acknowledgements}

We thank Guido Meier for helpful discussions. 
M.~T.~thanks M. Preuninger for support. 
We acknowledge support from the SFB 668 (Project B16) and from the SFB TR 12 of
the DFG.

\appendix*

\section{\label{app_deriv}Detailed calculations and magnetization profile}

Here we provide a detailed derivation of the spin torque quoted in the
main text and give the expressions valid for arbitrary Rashba coupling
$\alpha_R$.  In addition, we briefly discuss the space-time 
profile of the magnetization.

\subsection{Derivation of the spin torque}

Since we treat a nanowire in the one-dimensional (1D)
limit, we employ a Luttinger liquid (LL) description for the 
itinerant electrons. While there are several representations for the LL
Hamiltonian, we choose the Sugawara formulation which is most convenient here.
As discussed in textbooks \cite{Gogolin}, after linearization 
of the spectrum, the spin sector of the
 kinetic energy is encoded  in the universal spin Hamiltonian 
 $H_0=\sum_{p} \frac{v}{2}\int dx :\fJp\cdot\fJp:,$
where $:\dots:$ stands for normal ordering, $v$ is the spin velocity 
(which essentially equals the Fermi velocity),
and the chiral spin density operators are 
\[
\fJp (x)= \half : c^{\dagger}_{p\sigma}(x) \boldsymbol \sigma_{\sigma\sigma'}
c_{p\sigma'}(x):
\]
for left- and right-movers ($p=R,L=+,-$), respectively \cite{Gogolin,frojdh}.
The 1D fermion operators $c_{p,\sigma}(x)$ describe a $p$-mover 
with spin projection $\sigma$, and 
$\boldsymbol \sigma$ is the vector of spin Pauli matrices.
The spin density operator is then given by ${\bf s}={\bf J}_R+{\bf J}_L$.
Similarly, chiral density operators are introduced as 
$j_p = \sum_\sigma c_{p,\sigma}^\dagger c_{p,\sigma}^{}$,
with expectation value $\langle j_p\rangle=\rho_p$. The
particle density is $\rho_c=\sum_p\rho_p$, and the current 
density is given by $I_c=ev\sum_p p\rho_p$.
The above description also applies when Coulomb interactions 
are important \cite{Gogolin}.

Within the low-energy LL approach, particles are 
constrained to have 1D momenta close to $k_x\approx pk_F$ with 
Fermi wavenumber $k_F$.  
This fact allows us to simplify the standard two-dimensional Rashba
spin-orbit single-particle Hamiltonian, 
$\tilde\alpha_R(\sigma_y k_x -\sigma_x k_y)$ 
with Rashba coupling $\tilde\alpha_R$,
by effectively putting $k_x\to pk_F$ and $k_y\to 0$. 
With the dimensionless Rashba coupling
$\alpha_R=2\tilde\alpha_R k_F /\dsd$ and using the definition
of $\fJp$, we  arrive at the second-quantized form
\begin{equation}
 H_{\rm{SOI}}= \dsd\alpha_R\sum_p p \int\, dx\  \fJp\cdot \fey . 
\end{equation}
Adding the $sd$ Hamiltonian,
 $H_{\rm{sd}}=\dsd\int\, dx\ \fs\cdot\fn$, describing the exchange
interaction between 
the magnetization $\fn$ and the spin density $\fs=\fJ_R+\fJ_L$ 
of the itinerant electrons, we arrive at the Hamiltonian 
\begin{equation}\label{eqham}
H_{\rm el}= H_0+\dsd\sum_{p=\pm}\int dx  \ \fJp\cdot
( \fn+ p \alpha_R \mathbf e_{y} ).
\end{equation}
The spin torque, 
$\mathbf T = -\dsd \fn\times \langle\fs\rangle$, entering the LLG
equation can then be calculated by solving the Heisenberg equations of 
motion (EOM) for $\fJp$. With the standard summation convention,
these operators obey the Kac-Moody algebra \cite{Gogolin}
\begin{eqnarray*}
 \comm{J^a_p(x),J^b_{p'}(x')}&=& ip\delta^{ab}\delta^{pp'}\dx
\delta(x-x')\\ 
 && +i\delta^{pp'}\epsilon^{abc}J^c_p\delta(x-x')\, ,
\end{eqnarray*}
with the Kronecker symbol $\delta^{ab}$ and the Levi-Civita symbol 
$\epsilon^{abc}$. The Heisenberg EOM then reads
\[
\dt\fJp + pv\dx\fJp =  - \dsd \fJp\times (\fn+p\alpha_R\mathbf e_y),
\]
where a term $\propto\dx \fn$, irrelevant to the following discussion,
has been omitted.  Itinerant electron spins are 
also subject to relaxation processes not captured by the above
Hamiltonian, e.g., due to quasi-elastic or magnetic
disorder effects and/or additional spin-orbit coupling mechanisms.
In this work, we model relaxation on phenomenological grounds by adding
relaxation terms to the EOM:
\begin{eqnarray} \label{eqheom2}
\dt\fJp + pv\dx\fJp & = & - \dsd \fJp\times (\fn+p\alpha_R\mathbf e_y)\\
\nonumber &-& \sum_{\nu=i,ii}\beta^{(\nu)}\dsd \left(\fJp-
\fJp^{(\nu)}\right ) .
\end{eqnarray} 
The index $\nu$ stands for different relaxation channels. Each 
channel is characterized by
the rate $\beta^{(\nu)}\dsd$ and a corresponding 
quasi-stationary state $\fJp^{(\nu)}(x,t)$ that the system tries to reach.
We here include two relevant channels:
One ($\nu=i$) is provided by the externally imposed spin-polarized current
$I_s=\hbar P I_c/(2e)$, 
where a $p$-mover has density
$\rho_p$ and the spin polarization factor is $P$, with $|P|\le 1$.
This implies the stationary distribution 
${\bf J}_p^{(i)}=(P\rho_p/2) \fn$.
The second mechanism ($\nu=ii$) describes intrinsic relaxation to 
the stationary solution of Eq.~(\ref{eqheom2}), 
$\fJp^{(ii)}=(P\rho_p/2) [\fn+p\alpha_R\fey]$. 

In order to solve Eq.~(\ref{eqheom2}) for 
the physically relevant case of slow  space-time
variation of the magnetization, 
we now perform an iterative gradient expansion. In this approach\cite{mt2007},
$\fJp$ is expanded in orders of derivatives of the
magnetization unit vector $\fn(x,t)$,
$\fJp=\sum_{k=0}^{\infty}\fJp^{(k)}$,
where $\fJp^{(k)}$ depends on
space-time derivatives of $\fn$ of $k$th order only. 
With this, the EOM (\ref{eqheom2}) can be rearranged according to orders of $k$,
allowing for an iterative solution.  In this scheme, we first solve
the $k=0$ equations, then use this solution to obtain the $k=1$ term,
and so on. In particular, for $k=0$, we have to solve
\begin{eqnarray}
 0&=&-\dsd\fJp^{(0)}\times(\fn+p\alpha_R\mathbf e_y) - \\
\nonumber &&- \sum_{\nu=i,ii}\beta^{(\nu)}\dsd \left(\fJp^{(0)}-
\fJp^{(\nu)}\right ).
\end{eqnarray}
The $k$th-order contribution to the spin torque, $\mathbf T^{(k)}$,
then follows from
\[
 \mathbf T =\sum_{k=0}^{\infty}\mathbf T^{(k)}, \quad
 \mathbf T^{(k)} = -\dsd \fn\times \sum_p \fJp^{(k)} .
\]
For sufficiently smooth space-time variation of the magnetization,
the lowest few orders capture the relevant physics.  Due to 
the complexity of the higher-order terms, we here restrict
the calculations to zeroth and first order terms.

\subsection{Results for the spin torque}

As outlined above, we have ${\bf T}(x,t)={\bf T}^{(0)}+{\bf T}^{(1)}$,
where the first term involves no space-time derivatives $\partial_{x,t}\fn$
and the second contains exactly one first-order derivative.
Some algebra yields $\mathbf T^{(0)}=\mathbf T_1+\mathbf T_2$ with 
\begin{widetext}
\begin{eqnarray*}
\mathbf T_1&=&\frac{\dsd\alpha_R }{v\mathcal N }\left\{-I_s\left[1+\alpha_
R^2(1-2n_y^2)+\frac{\beta^{(ii)}}{\bi+\bii}
\alpha_R^2(1-2n_y^2+\alpha_R^2)\right] \right.\\
&&\left.+\rho_sv\frac{\beta^{(ii)}}{\bi+\bii}\alpha_Rn_y(\alpha_R^2-1)
\right\}\fn\times\mathbf e_y \, ,\\
\mathbf T_2&=&\bi\frac{\dsd\alpha_R}{v\mathcal N}\left[I_s(1+\alpha_R^2)+
2\rho_sv\alpha_Rn_y\right]\fn\times(\fn\times\mathbf e_y)\, ,\\
\mathcal N &=& (1+\ar^2)^2-4\ar^2 n_y^2\ ,
\end{eqnarray*}
where $\rho_s=P\rho_c/(2e)$. We stress that these results are valid
for arbitrary $\alpha_R$.
We also provide explicit expressions for the next order:
\begin{equation*}
\mathbf
T^{(1)}=\mathcal N^{-2}\sum_{\nu=x,t}
\left( A_{\nu}\partial_{\nu}\fn+B_{\nu}\fn\times\partial_{\nu}
\fn\right)-H^{(1)} _R\fn\times\mathbf e_y
\end{equation*}
with
\begin{eqnarray*}
(\bi+\bii) A_t&=& -\frac{I_s}{v}(\alpha_R^2-1)\alpha_R n_y \left\{\bii\mathcal
N-2\bi\left[1-\ar^4+2\ar^2(\ar^2-1)n_y^2\right]\right\} \\
&&+\rho_s\left\{-\bii\mathcal N \left[1+\alpha_R^2(1-2n_y^2)\right]
-\bi\left[\mathcal N+\alpha_R^2n_y(8\alpha_R^2(n_y^2-1)-\mathcal
N)\right]\right\}\, ,\\
(\bi+\bii) A_x&=& I_s\left\{-\bii\mathcal N \left[1+\alpha_R^2(1-2n_y^2)
\right]-\bi\left[\mathcal N+\alpha_R^2n_y(8\alpha_R^2(n_y^2-1)-\mathcal
N)\right]\right\}\\
&& - \rho_s v(\alpha_R^2-1)\alpha_R n_y\left\{\bii\mathcal N+2\bi
\left[1+\alpha_R^2(1-2n_y^2)\right]\right\}\, ,\\
B_t&=& \frac{I_s\ar n_y}{v}\left[4\bi(\ar^4-1)+2\bii\mathcal N\right]+
+\rho_s\mathcal N\left[\bii(1+\ar^2)+\bi(\ar^2-1)\right]\, ,\\
B_x &=& I_s\bi(1-\ar^2)(\mathcal N+8\ar^2 n_y^2)+4\rho_s v\bi\ar n_y(\ar^4-1)\, ,
\end{eqnarray*}
and 
\begin{eqnarray*}
 H_R^{(1)}&=&\mathcal N^{-2}\sum_{\nu=x,t}\left(h_{\nu}\partial_{\nu} 
n_y+h_{\nu}'\partial_{\nu}\fn\cdot(\fn\times\mathbf e_y)\right)\, ,\\
(\bi+\bii)^2 h_t&=&-\frac{I_s\ar^3 n_y}{v}\left\{2\bii\mathcal N+
\bi\left[-1+3\ar^4+2\ar^2(1-2n_y^2)\right]\right\}\\
&&-\rho_s\ar^2\left\{\bi\ar^2\left[(1+\ar^2)^2-4n_y^2\right]
+\bii(1+\ar^2)\mathcal N\right\}\, ,\\
(\bi+\bii)^2 h_x&=& -I_s\ar^2\left[\bii(1+\ar^2)\mathcal N+\bi\ar n_y\mathcal N\right]\\
&&+\rho_sv\ar^3 n_y\left\{2\bii\mathcal
N+\bi\left[-1+3\ar^4+2\ar^2(1-2n_y^2)\right]\right\}\, ,\\
(\bi+\bii) h_t' &=& -4\bi\frac{I_s}{v}n_y\ar^3(1+\ar^2)+\bi \rho_s\ar^2(\mathcal
N+8\ar^2 n_y^2)\, ,\\
(\bi+\bii) h_x' &=& \bi I_s\left[(\ar+\ar^3)^2+4\ar^2n_y^2\right]-4\bi \rho_s\ar^3 n_y(1+\ar^2)\ .
\end{eqnarray*}
\end{widetext}

\subsection{Time-dependent magnetization profile}

\begin{figure}[tb]
\centering
 \includegraphics[width=.98\linewidth]{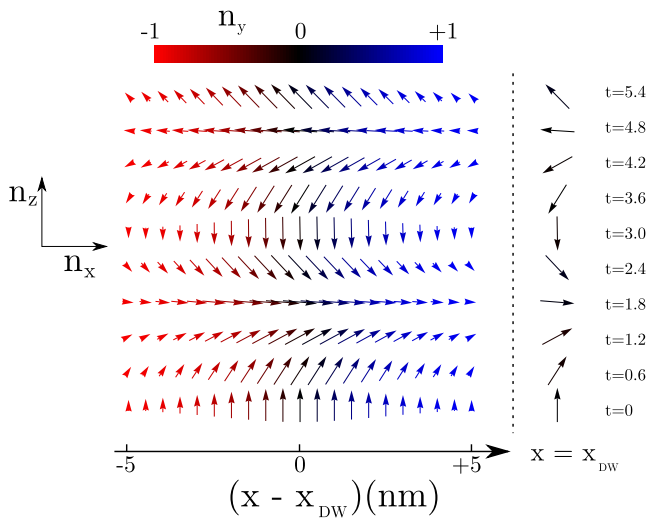}
\caption{\label{figure6}
(Color online) The magnetization profile in the
vicinity of the DW center $x\approx x_{DW}$ at different times $t$ (in
nanoseconds, from the bottom to the top). Arrows indicate $n_{x,z}(x-x_{DW},t)$,
while $n_y$ is color-coded. For clarity, we show the
magnetization profile right at the DW center $x=x_{DW}$ in a separate column on
the right of the profile. 
Due to the finite Rashba field, the magnetization precesses around the
$y$-axis, i.e., the direction of the Rashba field.
Parameters are as in Fig.~3 for constant current.
}
\end{figure}

When the Rashba field exceeds a distinct value,
the magnetization starts to precess around the axis 
defined by the field, in this case the $y$-axis. 
Let us illustrate such a scenario, see  Fig.\ \ref{figure6}.
As long as a spin-polarized current $I_s$ flows, a nonequilibrium 
Rashba field is created, which then causes precession of the magnetization 
$\fn$ around this field.  The resulting precession period equals  
two periods of the velocity variation,
 cf.~Fig.~3. When the current is switched off (not shown in
the figure), the DW relaxes slowly back to its stationary state, i.e., the
state for $t=0$ in the figure.

%
%
%
%

\end{document}